\documentclass{article}
\pdfoutput=1

\PassOptionsToPackage{numbers, compress}{natbib}
\usepackage[preprint]{neurips_2019}

\usepackage[utf8]{inputenc} 
\usepackage[T1]{fontenc}    
\usepackage{hyperref}       
\usepackage{url}            
\usepackage{booktabs}       
\usepackage{amsfonts}       
\usepackage{microtype}      
\usepackage{graphicx}
\usepackage{comment}
\usepackage{amsmath}
\usepackage{dirtytalk}
\usepackage[toc,page]{appendix}

\usepackage[linesnumbered,ruled,vlined]{algorithm2e}

\title{Roof Age Determination for the Automated Site-Selection of Rooftop Solar}

\author{
   Chris Heinrich \\
   Qrithm \\
   \And
   Michael Laskin \\ 
   UC Berkeley, Qrithm \\ 
   \And
   Simas Glinskis \\
   University of Chicago \\ 
   \And
   Evert van Nieuwenburg \\
   Caltech \\
}

\begin{document}

\maketitle
\begin{abstract}

Rooftop solar is one of the most promising tools for drawing down greenhouse gas (GHG) emissions and is cost-competitive with fossil fuels in many areas of the world today. One of the most important criteria for determining the suitability of a building for rooftop solar is the current age of its roof. The reason for this is simple -- because rooftop solar installations are long-lived, the roof needs to be new enough to last for the lifetime of the solar array or old enough to justify being replaced. In this paper we present a data-driven method for determining the age of a roof from historical satellite imagery, which removes one of the last obstacles to a fully automated pipeline for rooftop solar site selection. We estimate that a full solution to this problem would reduce customer acquisition costs for rooftop solar by $\sim$20\%, leading to an additional $\sim$750 megatons of CO$_2$ displaced between 2020 and 2050. 

\end{abstract}

\section{Introduction}

The last 10 years have seen a dramatic drop in the cost of electricity generated from solar photovoltaic (PV) systems \cite{fu2018us}. This drop in costs directly leads to an increased rate of deployment of PV systems, and therefore a reduction in future emissions of GHGs from fossil fuels. Once considered too expensive to be useful in the fight against climate change, ground mount and rooftop solar are now both ranked among the top 10 most promising methods for reducing GHGs. Rooftop solar alone is estimated to have the potential to offset 24.6 gigatons of CO$_2$ by 2050 if the fraction of global electricity produced by rooftop solar grows from today's level of .4$\%$  to 7$\%$ by 2050 \cite{hawken2017drawdown}. 

In order to reach or exceed this 7$\%$ level, however,  the cost of installing solar needs to drop even further. The hard costs of installing solar, including materials and equipment, have dropped to the point where now over two-thirds of the cost of installing solar comes from `soft costs' \cite{friedman2013benchmarking}\cite{sird} such as labor, permitting and customer acquisition costs. Customer acquisition costs (CAC) alone can account for up to 20\% of the \emph{total} costs of installing solar \cite{mond2017us} \cite{friedman2013benchmarking}. By developing automated methods for identifying and ranking potential solar installation sites,  solar developers can focus their sales and marketing budgets on the most promising sites, and thereby reduce their customer acquisition costs. In turn, this reduction in cost leads to an overall increased rate in the deployment of solar power \cite{sird}. Given that the roof of a building is easily visible from aerial images, global-scale satellite image data can be brought to bear on the site-selection problem, and this is precisely where machine learning (ML) can have a major impact. 

In this work we study how satellite imagery can be used to estimate the age of a building's roof. The roof age is a critical piece of information for solar developers because it directly impacts whether a building is currently well suited for rooftop solar. This is because rooftop solar installations last for 25+ years, and it is important that the roof does not need to be replaced during the lifetime of the solar array because the additional labor costs of removing and reinstalling the solar array while replacing the roof would typically cripple the economics of the solar project. This is especially true for commercial and industrial buildings where large, complex solar arrays typically cover all available roof area. Solar developers therefore generally seek out rooftops which are either 0-4 years or 25+ years of age, since in the first case the existing roof will likely last long enough, and in the latter case the roof is already old enough to justify replacement in conjunction with the installation of the solar array.

Although roof age can usually be determined by consulting the building's owner, making contact with the building owner, particularly for commercial and industrial buildings, can take considerable time and effort. This is where an automated estimation of roof age can lower customer acquisition costs, and therefore accelerate the deployment of rooftop solar \cite{sird}. We estimate that a full solution to this problem would displace an additional 750 megatons of CO$_2$ between the years of 2020 and 2050, with the details behind this estimated included in Appendix \ref{co2_savings}. 

\section{Related Work}\label{related-work}
The general site-selection problem for rooftop solar has multiple components, a number of which have been studied previously. One important problem is to estimate the potential size and output of a solar array on a given rooftop. Some pioneering approaches to automate this process include Mapdwell \cite{jakubiec2012towards},  as well as Project Sunroof \cite{projectsunroof}, which used LIDAR data to produce solar array size and shading estimates. More recently Ref. \cite{lee2019deeproof} proposed a data-driven method for estimating solar array size using only RGB satellite images and ML. Another component of the site-selection problem involves estimating a building owner's current electricity bill by estimating the energy consumption of the building, to which ML has been applied in Refs. \cite{kreider1995building} and \cite{amasyali2018review}. However, to the best of our knowledge, ours is the first work to study data-driven approaches to the roof age estimation problem in particular, and we believe a robust solution to this problem removes one of the last obstacles to a fully automated approach to rooftop solar site-selection.

\section{Methodology}
We propose to determine the age of a roof using only satellite imagery, which is widely available and global in scope. Instead of attempting to directly regress the age of a roof from a present-day satellite image, however, we instead propose to use historical satellite imagery to determine the year in which the roof was last replaced, thereby solving for the roof age as the number of years since that date. If no reroof is detected during the range of available historical satellite imagery, then the age of the roof is determined to be greater than that range.

\subsection{Data}\label{data}

\begin{figure}[!t]
  \centering
     \includegraphics[width=0.135\textwidth]{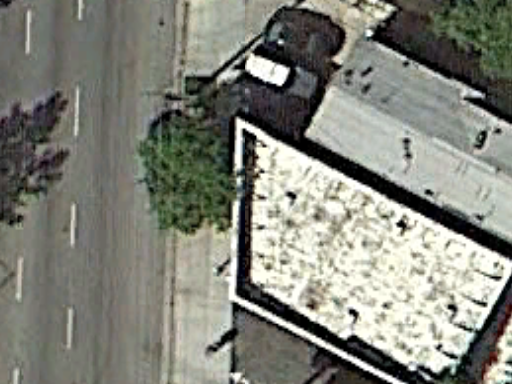}
   \includegraphics[width=0.135\textwidth]{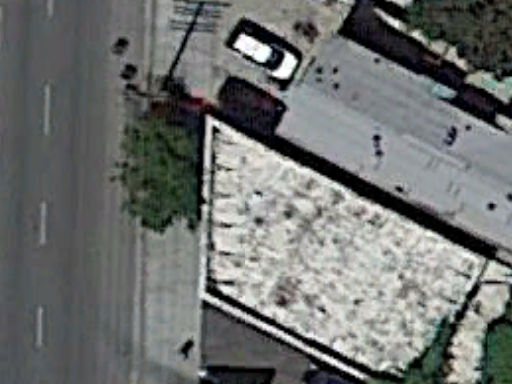}
 \includegraphics[width=0.135\textwidth]{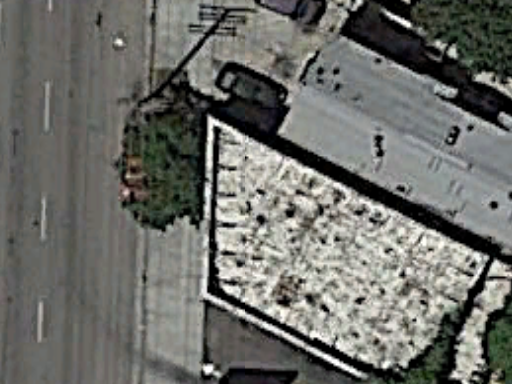}
  \includegraphics[width=0.135\textwidth]{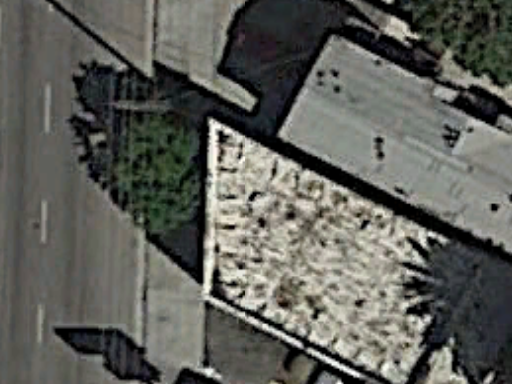}
  \includegraphics[width=0.135\textwidth]{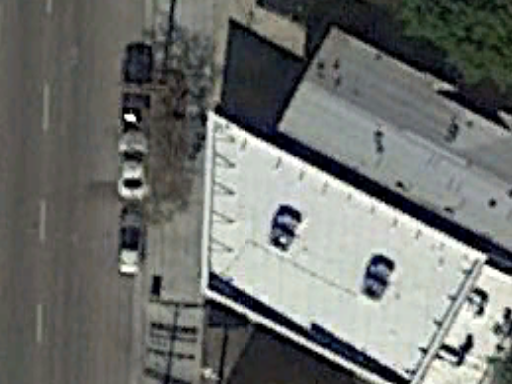}
  \includegraphics[width=0.135\textwidth]{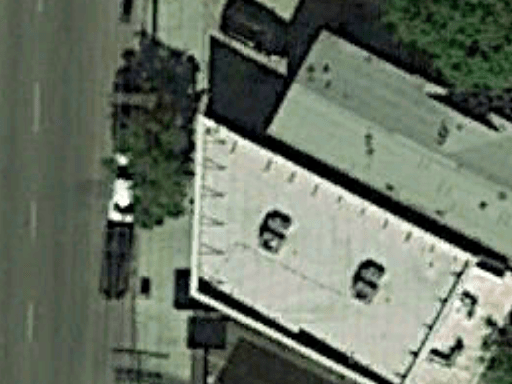}
  \includegraphics[width=0.135\textwidth]{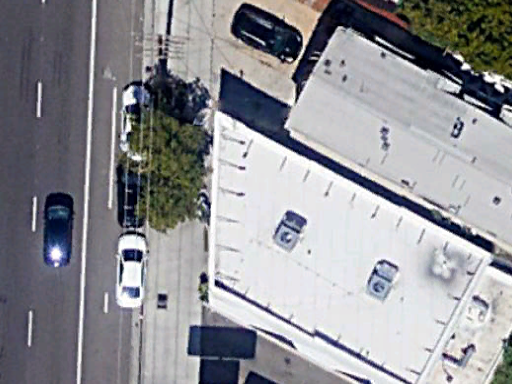}
    \includegraphics[width=0.135\textwidth]{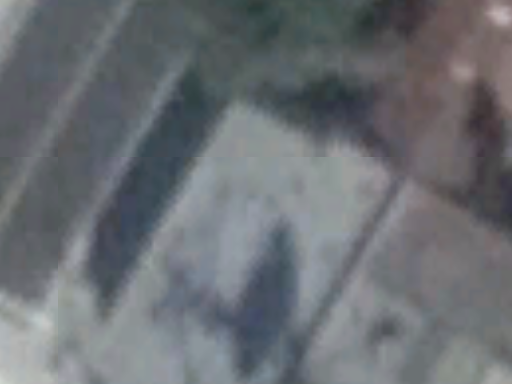}
  \includegraphics[width=0.135\textwidth]{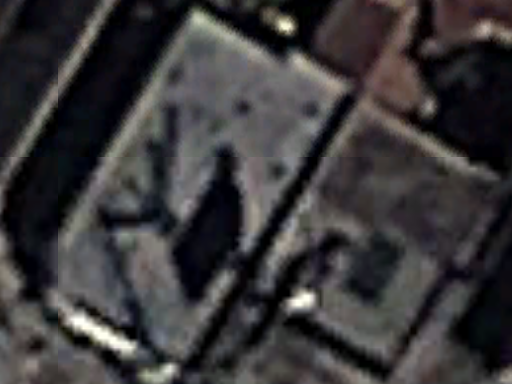}
   \includegraphics[width=0.135\textwidth]{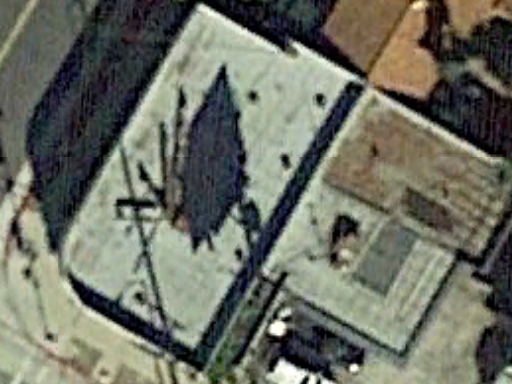}
  \includegraphics[width=0.135\textwidth]{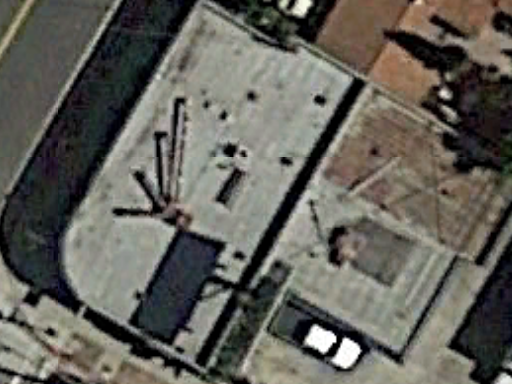}
  \includegraphics[width=0.135\textwidth]{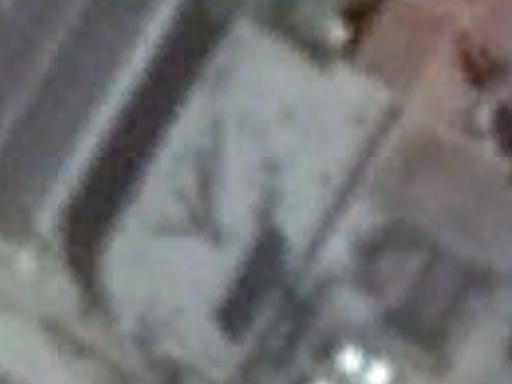}
  \includegraphics[width=0.135\textwidth]{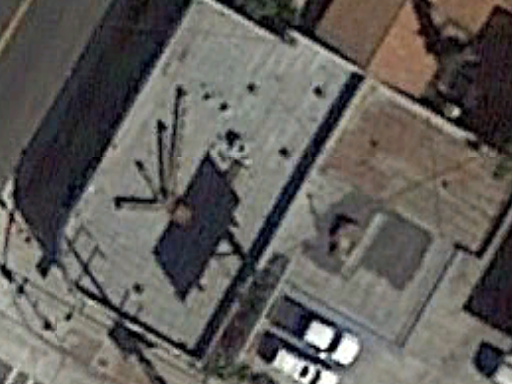}
  \includegraphics[width=0.135\textwidth]{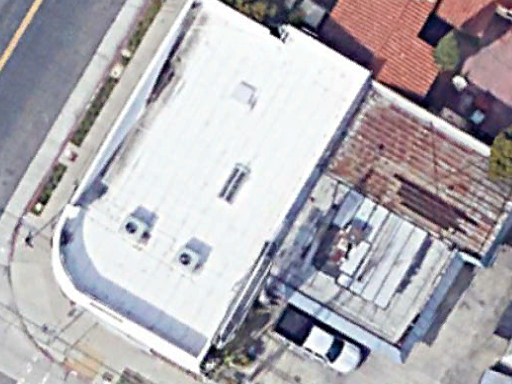}
  \caption{Example sequences of rooftop images taken in the years 2012-2018. The top row shows an easy example, with a clear roof change in the fifth column, meanwhile the second row shows a harder example which exhibits some of the challenges of this problem, including image blur and varying image exposures which can be mistaken for a changing roof. }
  \label{fig:roofs}
\end{figure}

We introduce a new dataset of commercial and industrial building roof images for studying this problem. The dataset consists of 1,610 images of building rooftops, covering over 230 buildings in Southern California. For each property we include one image per year in the range 2012-2018 and indicate (a) whether the roof was replaced during the 2012-2018 time frame and (b) the year when the reroof took place if there was one. Approximately 180 of the 230 buildings in the dataset underwent a reroof, and the reroof year was obtained from public building permit data from the City of Los Angeles and verified by eye \cite{ladata}. The dataset was partitioned into subsets for training, validation and testing, consisting of 150, 25 and 55 buildings respectively. Example image sequences are shown in Fig. \ref{fig:roofs}. \footnote{See https://github.com/cpheinrich/reroofdata for access to the reroof dataset}

\subsection{Algorithms}

We train a $\beta$-VAE \cite{betavae}, an unsupervised generative model, see appendix \ref{appendix:algorithms}, to embed satellite images into latent codes $x \rightarrow z$. In addition we also train a binary classifier, on the latent vectors of all image pairs for a given building, to predict whether the two images correspond to the same roof or a different roof. At inference time, latent vectors for every image in the sequence are generated, and adjacent pairs are classified with the binary classifier: $(z_{t-1},z_{t}) \rightarrow p_t$, with $p_t$ being the probability that the roofs at time $t-1$ and $t$ are different. If $p_t < 0.5 \text{ for all } t$, then no reroof is predicted for the image sequences, otherwise the transition year is determined to be $T = \text{argmax}_t( p_t)$. See Appendix \ref{hyperparams} for additional implementation details.

\section{Results}\label{results}
The metrics used to evaluate performance are (a) the reroof detection accuracy, i.e. the fraction of buildings for which it was correctly predicted whether there was or was not a reroof in the image sequence and (b) the average error in years, $\sum_i|y^i_{\textrm{true}}-y^i_{\textrm{pred}}| / N$. The average error metric is only computed on the buildings for which it was correctly determined that a reroof did take place. The average error is a useful metric because, for the problem at hand, it is generally sufficient to know the approximate roof age rather the exact year. 

We compare the $\beta$-VAE method to a categorical distribution baseline fit to the training data in Table \ref{table:results}. For the categorical baseline, a reroof year, or a no-reroof label, were randomly guessed according to the distribution of labels in the training dataset.  

\begin{table}
\centering
\begin{tabular}{lcc}
\hline
\textbf{Method} & \textbf{Reroof detection accuracy} & \textbf{Avg error (years)}   \\
\hline
$\beta$-VAE &  $\mathbf{0.872} $        &  $\mathbf{0.680}$     \\
Categorical baseline &  0.648         & 1.868            \\
\end{tabular}
  \caption{Evaluation results on the test dataset.}
  \label{table:results}
\end{table}

\subsection{Discussion}

We see from Table \ref{table:results} that the $\beta$-VAE significantly outperforms the categorical baseline on both metrics. While the historical range of this dataset is still limited, these results could already be used to suggest that buildings whose roof was replaced in 2012 or 2013 are less ideal targets for solar developers. In addition to the categorical baseline, we also compared the $\beta$-VAE to non-learning based methods that used features such as zero-normalized cross correlation, and normalized color intensity, to detect roof transitions but found that these alternative methods did not outperform the categorical baseline, providing additional justification for the use of learning based methods to adequately solve this problem. 

Several confounding factors make the problem of roof-age estimation from satellite images challenging. The quality and resolution of satellite image data can vary drastically year-to-year due to environmental factors, such as weather conditions and time of day, as well as sensor heterogeneity across data providers. We also found that wide-area high-resolution satellite imagery is less available before 2010, and virtually non-existent prior to the 1980s. Fortunately, the availability of high-resolution satellite images will only grow over time, and we expect our method to increase in quality and utility as more data becomes available. 

While building permit data can also be used to determine roof age, this data source comes with its own set of problems, including limited or difficult access, incomplete or false records, and non-standardized formatting across municipalities. The attractiveness of the satellite-image based approach is that it could be applied on a national, or even global scale, using only a single data source.

\section{Conclusion}
In this paper, we argued that automated roof-age estimation will enable faster large-scale deployment of rooftop solar and showed that it is possible to make such predictions within a reasonable margin of error using historical satellite imagery. We also introduced a new dataset to enable the continued study of data-driven approaches to this problem. Some interesting future directions for this work include developing methods that are more robust to confounding factors such as image blur and variation in satellite image sources, as well as applying the method to residential rooftops. It would also be useful to solve zero-shot roof-age estimation from a single image instead of detecting the year the roof was replaced.

\bibliographystyle{unsrt}
\bibliography{refs}

\begin{appendices}

\section{Algorithms}
\label{appendix:algorithms}
\subsection{Amortized Variational Inference}\label{vae-summary}

One unsupervised method for predicting roof age is amortized variational inference \cite{vae}. This method estimates the distribution of observed data $p(x)$ through a low dimensional latent distribution $p(z)$. The program encodes data into latent codes with a parametrized function $q_\theta (z|x)$ and decodes with a different parametrized function $p_\phi (x|z)$. The parameters of both functions are updated by maximizing the evidence lower bound (ELBO).

\begin{align}
     \mathbb{E}_{z \sim q_\theta} \left [\log p(x) \right ]   \geq   \mathbb{E}_{z \sim q_\theta} \left [ \log p_\phi(x|z) \right ] - \beta \mathbb{D}_{KL} \Big ( q_\theta(z|x) \big \Vert p(z) \Big ) 
\end{align}

A common choice for a prior over the latents is the unit Gaussian $p(z) = \mathcal N (0,I)$, and the posterior is also Gaussian $q_\theta (z|x) = \mathcal N (\mu,\sigma)$, where $\mu$ and $\sigma$ are learned by maximizing the ELBO forming maximum a posteriori probability estimate. When $\beta=1$, this algorithm is known as the Variational Autoencoder (VAE), and for $\beta > 1$ it is called a $\beta$-VAE \cite{betavae}.

\section{Hyperparameters}\label{hyperparams}

We preprocessed raw satellite images to $64 \times 64$ pixel images centered around the buildings' roofs and during training applied random transformations to brightness, contrast, and saturation. For the VAE architecture, we included four convolutional layers that downsampled the images and a fully connected dense layer that yielded 128 dimensional latent codes followed by a stack of three additional residual layers, such that the latent space had in total 128 dimensions.  During optimization we used Adam \cite{adam} with a learning rate of $\text{lr} = 3 \cdot 10^{-4}$. For the disentanglement parameter $\beta$ in the $\beta$-VAE we used $\beta=1$. The binary classifier had four fully connected layers with dropout that downsample the input 256-dimensional concatenated latent code in a one-dimensional logit passed through a sigmoid.  During optimization we used Adam with a learning rate of $\text{lr} = 1 \cdot 10^{-3}$.

\section{CO$_2$ offset calculation}\label{co2_savings}

In this section we estimate the potential impact of this work on GHG emissions, measured in terms of megatons of CO$_2$ offset between 2020 and 2050. While it is of course impossible to provide an exact number, we seek to provide a defensible order of magnitude estimate by modeling the different aspects of the problem and making reasonable assumptions.  This estimation problem can be broken down into two parts:
\begin{enumerate}
    \item Estimate how prior knowledge of the building roof age can reduce customer acquisition costs for solar developers.
    \item Estimate how much this reduction in solar installation cost will increase the rate of deployment of rooftop solar.
\end{enumerate}

\subsection{Impact of roof age knowledge on CAC}
Customer acquisition costs account for 10-20\% of the total system cost of installed solar. Broadly speaking, a simplified model of the customer acquisition pipeline can be represented as a funnel with two parts, with the top of the funnel consisting of outbound sales and marketing, and the bottom of the funnel consisting of working with the potential client to collect more detailed information about the building, the project and ultimately getting the client to sign a contract. Transition from the top of the funnel to the bottom of the funnel is generally triggered by a key event in the early stage of the sales process that qualifies the lead as viable, such as the collection of a utility bill or the signing of a letter of intent to purchase solar. Prior knowledge of the roof age will reduce top of funnel customer acquisition expenses, but will not generally reduce the bottom of funnel expenses because at this stage the roof age is either known, or because direct contact with the client is established, can easily be retrieved. While the ratio of top-of-funnel expenses to bottom of funnel expenses will vary due to a number of reasons, a reasonable estimate for this ratio is 40\%. Without any \emph{a priori} knowledge of roof age, we will assume that the distribution of building roof ages is uniform in the range $[0,39]$ years. With this assumption, we can see that only half of the potential buildings would have a roof in the viable age range of $[0,4]\cup [25,39]$, whereas the rest of the roofs would fall in the \emph{dead zone}. This figure of 50\% also agrees with actual figures the authors have observed when working with solar developers to conduct data-driven outbound marketing campaigns. With this in mind, we see that 50\% of top-of-funnel sales and marketing budget is spent on buildings that have invalid roof ages, and if these ages were known from the start, the total customer acquisition costs could be reduced by $\sim20\%$.

\subsection{Impact of CAC reduction on CO$_2$ displacement}
Assuming CAC represents 10\% of the total costs of solar installation \footnote{This figure may be conservative, some recent estimates put CAC at up to 20\% of installation costs~\cite{mond2017us}.}, the aforementioned reduction of 20\% of CAC results in an overall savings of $\sim$2\%. These savings can be directly translated into an estimated decrease of CO$_2$ emissions through an expected increase rate of solar deployment. For small changes in cost such as this one, we estimate the rate of deployment and the total capacity from solar installations to increase \emph{linearly} with a decrease of costs, in accordance with observed trends \cite{sird}. The total capacity of electricity generated from rooftop solar should therefore increase by $\sim 2\%$, meaning we can reduce the electricity generated from natural gas by an amount equivalent to the that generated by this increase in solar capacity. This 2\% increase in solar would generate roughly 1200 thousand MW, equalling 0.1\% of the energy produced by natural gas. Whilst this small percentage may not seem like much, it is responsible for 25 megatons of CO$_2$ per year. Over the 30 year period of 2020 -- 2050, this would amount to an additional displacement of about 750 megaton of CO$_2$ on top of the business-as-usual prediction of 24.6 gigatons \cite{hawken2017drawdown}.

\end{appendices}

\end{document}